
\magnification=1200
\font\tit=cmssbx10 scaled \magstep2
\hsize=16.2truecm \vsize=23.5truecm
\nopagenumbers
\footline={\ifnum\pageno>0\hfil\rm\number\pageno\hfil\fi}
\def\cv{\hfil\break}
\def\Re{{\rm Re}} \def\Im{{\rm Im}}
\def\cldot{\!\cdot\!}
\def\tr#1{{\bf #1}}
\def\*{\star}
\def\ac{\alpha '_c}
\def\hf{{\textstyle {1\over 2}}}
\def\G{\Gamma}
\def\~#1{\mathaccent "7E #1}
\def\=#1{\mathaccent 22 #1}
\def\^#1{\mathaccent 94 #1}
\def\q{{\bf q}}
\def\k{{\bf k}}

\def\Q{{{\bf q}^2}}
\def\a{{\rm a}}
\def\b{{\rm b}}
\def\A{{\rm A}}
\def\C{{\rm B}}
\def\gsim{\mathrel{\rlap{\lower0.5ex\hbox{$\sim$}}{\raise0.5ex\hbox{$>$}}}}
\def\refno#1{\item{[#1]}}
\def\sez#1{\bigbreak\vskip15truemm plus 5truemm minus 1truemm
 \leftline{\bf #1}\nobreak\bigskip}
\def\arrow#1{\hbox to #1truemm{\rightarrowfill}}
\def\limrel#1\for #2{\mathrel{\mathop{\kern0pt#1}\limits_{#2}}}
\def\fig#1 {\item{\bf Fig. #1.~}}
\pageno=0
\baselineskip=15pt
\null\vskip-1.5truecm\rightline{NORDITA 92/45}
\vskip3mm\rightline{July, 1992}
\vglue 2truecm
\centerline{\tit Superstring One-Loop and Gravitino}
\vskip3mm\centerline{\tit Contributions to Planckian Scattering}

\vglue 1.5truecm
\centerline{Alessandro Bellini\footnote{*}{Supported in part by
A. Della Riccia Foundation.}}

\centerline{\it NORDITA, Blegdamsvej 17, DK-2100 Copenhagen \O, Denmark}

\centerline{\it and INFN, Sezione di Firenze, Largo E. Fermi 2, I-50125
Firenze,
Italy}
\bigskip
\centerline{Marco Ademollo and Marcello Ciafaloni}

\centerline {\it Dipartimento di Fisica dell'Universit\`a, Firenze and INFN,
Sezione di Firenze}

\centerline{\it Largo E. Fermi 2, I-50125 Firenze, Italy}

\vfil
\centerline {\bf Abstract}
\bigskip
Corrections to the semiclassical approximation in nearly forward Planckian
energy collisions are here reconsidered. Starting from the one-loop
superstring amplitude, we are able to disentangle the first subleading
high-energy contribution at large impact parameters, and we thus
directly compute the one-loop correction to the superstring eikonal.
By comparing this result with previous ones by Amati, Ciafaloni and
Veneziano (ACV) for pure gravity, we identify one-loop gravitino
contributions which agree with previous results by Lipatov. We finally
argue, on the basis of analyticity and unitarity, that gravitinos do
not contribute at all to the large distance two-loop ACV correction,
which thus acquires a universal ``classical'' interpretation.

\vglue1cm\break
\sez{1. Introduction}

\baselineskip=18pt
Gravitational scattering at Planckian energies and small angles was
investigated in the past [1-4] and is still under discussion [5,6],
in order to understand the role of short distances in string theory and
quantum gravity. A similar investigation was also performed [7] from
the point of view of reggeized graviton exchange in the multi-Regge
kinematics.

The high energy regime $2E=\sqrt s>m_{Planck}$ is characterized
[3] by a strong effective coupling $\alpha_G =Gs/\hbar$, so that
a resummation of the perturbative series is needed. Various approaches have
shown that the leading contributions to the scattering amplitude at a given
impact parameter $b$ yields a semiclassical, eikonal approximation
$$\eqalign {S\left( b,E\right) &=e^{2i\delta_0 (b,E)}
\,,\cr
\delta_0 &=-{Gs\over\hbar }\log b\cr}\eqno (1.1)$$
where the unobservable (IR singular) Coulomb phase has been omitted.

Corrections to the leading result (1.1) involve additional powers of $G$,
occurring in the dimensionless combinations
$$\eqalignno{{\hbar G\over b^2}&={\lambda_P^2\over b^2}\,,\qquad\qquad
\left(\lambda_P^2\equiv G\hbar\right)&(1.2\a )\cr
{G^2 s\over b^2}&={R^2\over b^2}=\alpha_G {\lambda_P^2\over b^2}\,,
&(1.2\b )\cr}$$
and, for the string case [3], also in the combination
$${G\hbar\over b^2}{\alpha '\hbar\over b^2}={\lambda_P^2\over b^2}
{\lambda_s^2\over b^2}\,,\qquad\qquad
\left(\lambda_s^2\equiv \alpha '\hbar\right).\eqno (1.3)$$

They are small for sufficiently large values of $b$ with respect to the Planck
length $\lambda_P$, the gravitational radius $R=2GE$, and, in the string case,
also the string length $\lambda_s$. Since $R$ is a classical parameter, we
refer to the corrections of type (1.2b) as classical ones, while we refer to
those of type in eq. (1.2a) (eq. (1.3)) as quantum (string) corrections.

No complete theory of such corrections exists so far. However, a large class
of string corrections (1.3) and the first perturbative terms of type (1.2)
were computed in ACV I [3] and ACV II [4]. In the $\alpha'\to 0$ limit their
result can be expressed in terms of an effective eikonal representation for
the elastic scattering amplitude
$$S(b,E)=\exp 2i\left(\delta_0+\delta_1+\delta_2\right)\,,\eqno (1.4)$$
where, for pure gravity,
$$\delta_1={6\over\pi}{G^2 s\over b^2}\log s\,,\qquad\qquad
\Re\delta_2=2 {G^3 s^2\over\hbar b^2}.\eqno (1.5)$$

The result (1.5) involve precisely the parameters
$\delta_1/\delta_0\simeq\lambda_P^2/ b^2$ and
$\delta_2/\delta_0\simeq R^2/b^2$ of eqs. (1.2), and were obtained
in ref. [4] by computing $\Im\delta_2$ directly, and using an analyticity
argument in order to obtain $\delta_1$ and $\Re\,\delta_2$ for the case of
pure gravity.

In this paper we reconsider the calculation of $\delta_1$ in superstring
theory,
by a direct evaluation of the high energy behavior of the one-loop string
amplitude.

Since the one-loop leading behavior was already analyzed in the past [3,8]
and is embodied in eq. (1.1) through the $O(G^2)$ term $i\delta_0^2$, the real
problem is to disentangle the first subleading contribution. Previous attempts
[8] have failed, due to the difficulty of properly subtracting the
leading term.

{}From the point of view of graviton Regge pole exchange, the relevant
contributions to the amplitude in momentum transfer \tr q space
$${1\over s}A (s,\Q)= 4\int d^2\tr b\, e^{i\tr b\cdot\tr q}{e^{2i\delta
(\tr b,E)}\over 2i}\eqno (1.6)$$
are the following:

(1) At tree level, corresponding to single Regge pole exchange%
\footnote{$^{\*}$}{We omit the shrinkage terms in the $\alpha'\to 0$
limit.}, we obtain
$${1\over s}\A^{(0)}(s,\Q)\simeq 4\~\delta_0 ={8\pi Gs\over\Q}\eqno (1.7)$$

(2) At one loop level we have similarly
$${1\over 4s}\A^{(1)}(s,\Q)=i\~\delta_0^2+\~\delta_1\,,\eqno (1.8)$$
where $\~\delta_n\equiv\~\delta_n(\tr q,E)$ indicate the Fourier transforms
of $\delta_n(\tr b,E)$. In particular
$$ i\~\delta_0^2\simeq i{2\pi G^2s^2\over \Q}\log\Q\eqno (1.9)$$
corresponds to the leading Regge cut, while
$$\~\delta_1\simeq -6G^2s^2\log s\log\Q\eqno (1.10)$$
is the Regge pole renormalization. The latter was already computed, on the
basis
of a generalized soft graviton emission current [7,9] in the model of reggeized
graviton exchange of ref. [7].

In order to disentangle the subleading contribution (1.10) in string theory,
we start considering the representation of the one-loop superstring amplitude
over the torus and we identify the region in moduli space which is relevant
for the pole renormalization (sect. 2). This region overlaps with the cut
region, and a careful subtraction is needed in order to obtain a finite
expression (sect. 3).

Our final result (sect. 4) has the same form as the one in eq. (1.5) for pure
gravity, but contains additional contributions that we ascribe to gravitinos,
as expected in SST theory, leading to N=8 supergravity in the field theory
limit. This interpretation agrees with Lipatov's [7] calculation of the
gravitino contributions.
\sez{2. High energy behavior of superstring one-loop amplitude}

The large $s$ and fixed $t$ asymptotic limit of the graviton-graviton
scattering amplitude was discussed in ACV I for the case of Type
II superstring theory to one loop. In that paper it was shown that the
leading contribution is given by the exchange of two gravireggeon poles,
corresponding to a Regge cut in the angular momentum $J$ plane. We give
in the following an expression containing the subleading term, which is
proved to be a double pole in the $J$-plane, according to the general
expectation of graviton Reggeization.

In general we expect quantum and
string corrections to contribute to this double pole at fixed $t$, but
in $D=4$ and in the limit $\ac\Q\to 0$ we shall show that only quantum
corrections of type (1.2a) survive. In principle this could be directly
achieved by Feynman graph computation in $N=8$ supergravity, the low energy
quantum field theory derived from Type II superstring with toroidal
compactification of six dimensions. However, the string calculation already
contains the UV-cutoff $\ac$, and turns out to be simpler.

The superstring scattering amplitude, in units
$\ac ={1\over 2}\alpha'= \hbar=c=1$, is given by
$$\eqalignno{A_4=-2g_D^2K_{cl}&\biggl\{{\G(-\hf s)\G(-\hf u)\G(-\hf t)
\over \G(1+\hf s)\G(1+\hf u)\G(1+\hf t)}\cr
&-g_{10}{1\over 2\left(16\pi^3\right)^2}
\int_{F} {d^2\tau\over (\Im\tau)^5} F_{2}(\tau,R_{c})\int
\prod_{r=a,b,c}d^2\nu_r \prod_{r < s} \chi_{rs}^{2k_r k_s}\biggr\}
&(2.1)\cr}$$
where $K_{cl}$ is the standard superstring kinematical factor
whose asymptotic behavior is
$$K_{cl}\sim \left(s\over 2\right)^4 \varepsilon_a\cldot\varepsilon_d\,
\varepsilon_b\cldot\varepsilon_c.$$
The integral in moduli space is over the fundamental region $F$ of the torus
topology [10], while $F_2$ is the toroidal compactification factor for closed
string. Finally $\chi_{rs}=\chi\bigl(e^{2\pi i(\nu_r-\nu_s)}, e^{2\pi
i\tau}\bigr)$ is
given in terms of the Jacobi $\Theta$ function and its asymptotic behavior
($\tau\to i\infty,\ {\nu/\tau} >0$ fixed) relevant for our calculation is
$$\chi_{rs}=\chi\left(e^{2\pi i(\nu_r-\nu_s)},e^{2\pi i\tau}\right)={\rm
const.}
\times\exp\biggl[-\pi {(\Im\nu)^2\over \Im\tau}+\Re\bigl(i\pi\nu+e^{2\pi i\nu}+
e^{2\pi i(\tau-\nu)}\bigr)\biggr]\eqno (2.2)$$

The $s$ and $t$ dependence in the integrand of (2.1) appears in the factor
$$ f=\left(\chi_{ad}\chi_{bc}\over\chi_{ac}\chi_{bd}\right)^{-t}
\left(\chi_{ab}\chi_{cd}\over\chi_{ac}\chi_{bd}\right)^{-s}
\eqno (2.3)$$
By using for the $\chi$'s the asymptotic form (2.2) we obtain
$${\chi_{ad}\chi_{bc}\over\chi_{ac}\chi_{bd}}
=\Bigl\vert 4\sin\pi(\nu_d-\nu_a)\sin\pi(\nu_b-\nu_c)
e^{-2\pi x(1-x)\tau_2}\Bigr\vert\,,\eqno (2.4\a )$$
where we have set \ $x=\Im (\nu_b+\nu_c)/2\tau_2\,$, \ $\tau_2=\Im\tau$, and
$$\eqalignno{&{\chi_{ab}\chi_{cd}\over\chi_{ac}\chi_{bd}}
=\exp\biggl\{{2\pi\over\tau_2}\Im (\nu_a-\nu_d)\Im (\nu_b-\nu_c)\cr
&\qquad -\Re\bigl[8\sin\pi(\nu_d-\nu_a)\sin\pi(\nu_b-\nu_c)e^{i\pi\tau}
\cos \pi(\tau-\nu_d-\nu_a+\nu_b+\nu_c)\bigr]\biggr\}.&(2.4\b )\cr }$$
In the limit $s\to i\infty$ there are two regions that
can contribute: one is the cut region discussed in ACV I
$$\Im (\nu_a-\nu_d)\Im (\nu_b-\nu_c)\sim O(s^{-1})\,,\qquad\quad
\tau_2\gsim O(\log s)\eqno (2.5\a )$$
the other is related to the pinching configuration of fig. 1, i.e.
$$ \vert\nu_d-\nu_a\vert\,\vert\nu_b-\nu_c\vert\sim O(s^{-1})\,,
\quad\qquad {\rm any}~~\tau_2.\eqno (2.5\b ) $$

The latter is the double pole region we are interested in, to which in general
all values of $\tau_2$ contribute. However, in the region $\Q\to 0$ relevant
for large values of $b$,
the dominant behavior will come from large values of $\tau_2$. This explains
a posteriori why the large $\tau_2$ behavior (2.2) is relevant, but also shows
that cut and pole regions do overlap, which makes it difficult to extract the
double pole correction.

In order to perform the cut subtraction we first recast eq.(2.1) in a more
manegeable form, valid in both regions (2.5a) and (2.5b), for all values of
the phases of $(\nu_d-\nu_a)$ and $(\nu_b-\nu_c)$.
Following ACV we introduce the variables $ a,b,c$ and perform the
integrations over $\Re \tau $ and $a$, which provide each a Bessel function.
This is still possible also in the region (2.5b) because the last term in the
exponent of (2.4b) can be written as:
$$\eqalignno {...&
+4\rho_+\rho_-\exp[-2\pi\tau_2 x -\pi\Im (\nu_d-\nu_a)]\cos \phi_1\cr
&+4\rho_+\rho_-\exp[-2\pi\tau_2 (1-x)+
\pi\Im (\nu_d-\nu_a)]\cos \phi_2 &(2.6)\cr}$$
where we have set $\nu_d=\tau$, we have introduced the shorthand notation
$$\rho_+=\vert\sin\pi(\nu_d-\nu_a)\vert\,, \qquad\qquad
\rho_-=\vert\sin\pi(\nu_b-\nu_c)\vert $$
and the two phases $\phi_{1,2}$ are linear in $\tau_1$ and $a$ and defined by
$$\phi_1=2\pi\tau_1+2\alpha_{bc}-\phi_2,\quad
\quad\quad\phi_2=2\pi a+\alpha_{bc}+\alpha_{ad},$$
where e.g. $\alpha_{bc}$ is
$$\alpha_{bc}=\arcsin {2\sinh\pi\Im (\nu_b-\nu_c)\cos\pi\Re (\nu_b-\nu_c)\over
\left[2\cosh 2\pi\Im (\nu_b-\nu_c)-2\cos 2\pi\Re (\nu_b-\nu_c)\right]^{1/2}.}$$

Finally, neglecting $\Im (\nu_d-\nu_a)$, with respect to $\tau_2 x$ in the
exponents of eq.(2.6), we obtain the following expression, which contains the
Regge cut and the double pole singularity:
$$ A_4=(8\pi G)^2\left(\sigma\over 2\right)^3 I,\qquad\qquad  s=i\sigma,
\quad\epsilon={4-D\over 2}\eqno (2.7\a )$$
$$\eqalignno{
I=&\int_0^1dx\int_0^{\infty}d\tau_2 \,(8\pi^2\tau_2 )^\epsilon
\int_0^{\tau_2}d\Im (\nu_d-\nu_a)\int_{-\tau_2}^{+\tau_2}d\Im (\nu_b-\nu_c)\cr
&\times {2\sigma\over\tau_2}\exp\biggl[i{2\pi\sigma\over\tau_2}
\Im (\nu_d-\nu_a)\,\Im (\nu_b-\nu_c)\biggr]\int_{-1/2}^{+1/2}\!d\Re
(\nu_d-\nu_a)\cr
&\times\int_{-1/2}^{+1/2}\!d\Re (\nu_b-\nu_c)\,(4\rho_+\rho_-)^{q^2}
\exp\bigl[-2\pi\tau_2 q^2 x(1-x)\bigr]\cr
&\times J_0\bigl(4\sigma\rho_+\rho_-e^{-2\pi\tau_2 x}\bigr)
J_0\bigl(4\sigma\rho_+\rho_-e^{-2\pi\tau_2 (1-x)}\bigr)
&(2.7\b )\cr}$$

The Regge cut dominates for $\Re\,t<2$. Such a dominant
contribution has to be carefully subtracted before we can calculate the
trajectory renormalization.
\sez{3. Cut subtraction and pole renormalization}

We now proceed to integrate the external leg insertions on two small circular
regions \ $|\nu_d-\nu_a|<c,\ \ |\nu_b-\nu_c|<c$ \
around the points $\nu_d=\nu_a$ and $\nu_b=\nu_c$. This pinching
region may be singled out alternatively, by studying the asymptotic behavior
of the integral expression in eq. (2.1) for $s\to i\infty$ with the method
of stationary phase, as first pointed out by Sundborg [8].
At the points above the phase is stationary in $\tau_2$ and $x$, and the
argument of the exponential in eq. (2.4b) is small, implying the
absence of non-Regge exponential dependence on $s$.

To do the integral, we introduce polar coordinates in both circular regions,
we perform the angular integrations by using the representation
$$ \int_0^{2\pi}{ d\theta_1d\theta_2\over 4\pi^2}\exp(
i2z\sin\theta_1\sin\theta_2)=J_0^2(z)$$
and we obtain the result
$$\eqalignno{
I&=\int_0^1dx\int_0^{\infty}d\tau_2 \,(8\pi^2\tau_2 )^\epsilon\,
e^{-2\pi\tau_2 q^2x(1-x)}\int d\rho_+d\rho_-\,
(4\rho_+\rho_-)^{q^2}{4\sigma\rho_+\rho_-\over \pi^2\tau_2}\cr
&\phantom{=}\times
J_0^2\Bigl({\sigma\rho_+\rho_-\over\pi\tau_2}\Bigr)
J_0\bigl(4\sigma\rho_+\rho_-e^{-2\pi\tau_2 x}\bigr)
J_0\bigl(4\sigma\rho_+\rho_-e^{-2\pi\tau_2 (1-x)}\bigr).
&(3.1)\cr}$$

We now expand the two last $J_0$ functions in power series of their arguments
and we replace the sums with Sommerfeld-Watson transforms in the $n,m$ complex
planes obtaining:
$$\eqalignno {
I&={(4\pi)^{\epsilon}\over 2\pi^3}\int_{\gamma}
{dn\over 2\pi i}{dm\over 2\pi i}
{\Gamma(-n)\Gamma(-m)\over \Gamma(n+1)\Gamma(m+1)}
\left(\sigma\over 2\right)^{2n+2m}\int_0^1 dx\int_0^\infty d\tau_2\,
{\tau_2}^{\epsilon} e^{-B\tau_2}\cr
&\phantom {=}\times
\int_0^c {d\rho_-d\rho_+} (\rho_-\rho_+)^{2n+2m+q^2}
\left[ \left({\sigma\over\tau_2}\rho_-\rho_+
J_0^2\Bigl({\sigma\over\pi\tau_2}\rho_-\rho_+\Bigr)-1\right)+1\right].
&(3.2\a )\cr}$$
where we have defined
$$B={\bf q}^2x(1-x)+2nx+2m(1-x)\eqno (3.2\b )$$

It is worthwhile to discuss more the expression (3.2a). We expect that the
magnitude of the two circular regions, in the pinched geometry of fig. 1,
is not relevant to asymptotics. This is certainly true for the integral of
the terms in round brackets, which is convergent for large $\sigma $ due to
the property
$$\sigma pJ_0^2\left({\sigma\over\pi}p\right) \limrel\arrow {10}\for
{\sigma\to\infty} 1+\sin\left(2{\sigma\over\pi}p\right)$$
but not for the last term in square brackets, which produces a factor
$\sim (c\sigma)^{2n+2m}$, to be interpreted as the tail of the cut
contribution in the pole region. By subtracting the last term we obtain the
pole contribution
$$\eqalignno {
I_{{\rm pole}}&={(4\pi)^{\epsilon}\over 2\pi^3}
\left(\sigma\over 2\right)^{-1-\Q}\log \sigma
\int_{\gamma}{dn\over 2\pi i}{dm\over 2\pi i}
{\Gamma(-n)\Gamma(-m)\over \Gamma(n+1)\Gamma(m+1)}\cr
&\phantom {=}
\times\int_0^1 dx\int_0^\infty d\tau_2\,{\tau_2}^{\epsilon} e^{-B\tau_2}
\int_0^\infty dz\, z^A\left[{2z\over\tau_2}J_0^2\left({2z\over\pi\tau_2}
\right)-1\right]\,,&(3.3\a )\cr}$$
where
$$A=2n+2m+\Q \eqno (3.3\b )$$
and we have kept only the term proportional to $\log \sigma$, to be
interpreted as a trajectory renormalization (double pole in the J-plane).
The integral with respect to $z$ in the above expression is well defined
for $A>-1$, and in particular in the region $A\sim 0$ we are interested in,
as a result of the cut subtraction. We evaluate this integral using an
alternative integral representation of the same analytic function valid
for $-2< A <-1$,
$$\int_0^\infty dz\, z^{A+1}J_0^2\left({z\over\pi}\right)=
\biggl({1\over 2\pi}\biggr)^{-A-1}\pi{\G (-A-1)\G (1+\hf A)
\over\G(-\hf A)^3}, $$
so we have finally\footnote{$^{\*}$}{The singularity at $A=-1$ of the
pole expression (3.4) is actually spurious,
because it cancels out with the cut term in the complete expression (3.2).}

$$\eqalignno {
I_{{\rm pole}}&={(4)^{\epsilon}\over 2\pi^4}
\left(\sigma\over 2\right)^{-1-\Q}\log \sigma
\int_{\gamma}{dn\over 2\pi i}{dm\over 2\pi i}
{\Gamma(-n)\Gamma(-m)\over \Gamma(n+1)\Gamma(m+1)}\cr
&\qquad\times
\int_0^1 dx\, {\left(\pi\over B\right)}^{\epsilon +A+2}
\G(\epsilon +A+2)
\pi{\G (-A-1)\G (1+\hf A)\over\G^3(-\hf A)}.&(3.4)\cr}$$

The $m$ and $n$ integrations of eq. (3.4) are performed in detail in
Appendix A in the $\ac {\q ^2}\to 0$ limit, by replacing the
integration in $x$ with a momentum integral in transverse space as in ACV I.
Here we only notice that the $\Q =0$ singularities in (3.4) come from
$B=0$ (cfr. eq. (3.2b)) and thus from the large $\tau_2$ region, as
mentioned before.

The final result, coming from eq. (3.4) and Appendix A, yields
the double pole amplitude
$${1\over s}\A_{{\rm pole}}^{(1)}(s,\Q)={8\pi G\over \Q}\log s\,
\delta\alpha (\Q)\,s^{1-\Q},\eqno (3.5)$$
corresponding to the J-plane projection
$$ A_J^{(1)}={1\over (J-2+{\bf q}^2)^2}{8\pi G\over {\bf q}^2}
\delta\alpha (\Q)\,,\eqno (3.6)$$
where $\delta\alpha $ is the trajectory renormalization
(for small $\ac\Q$ )
$$\delta\alpha=G\Q\int {d^2 k\over (2\pi)^2}\biggl[{(\Q-t_1-t_2)^2
\over t_1^2 t_2}+2{\Q -t_1 -t_2\over t_1t_2}+
(1\leftrightarrow 2)\biggr],\eqno (3.7\a )$$
$$t_1={\bf k}^2,\qquad\qquad t_2=({\bf q}-{\bf k})^2.\eqno (3.7\b )$$
Note that the singular $\Q\to 0$ behavior of this expression is not modified by
the following rewriting
$$\eqalignno{\delta\alpha &= 8\pi G {\q ^2\over {(2\pi)^3}}\int {d^2\k\over
\k ^2(\q -\k )^2}\biggl[ [\k\cdot (\q -\k )]^2\biggl({1\over \k ^2}+{1\over
(\q -\k )^2}\biggr)\cr
&\qquad \qquad\qquad
+4\k\cdot (\q -\k ) -\Q\biggr],&(3.8)\cr}$$
which coincides with the Lipatov result [7] for N=8 supergravity.
\sez{4. One loop correction to the eikonal and gravitino contributions}

The first subleading contribution to the one-loop string amplitude obtained in
eq. (3.5) is to be compared with the eikonal representation (1.6). Since the
leading Regge cut corresponds to the $i\~\delta_0^2$ term in eq. (1.8), the
result (3.5) yields directly the expression for $\delta_1^s$, i.e., the
superstring one loop correction to the eikonal
$$\delta_1^s(b,s)={2\over\pi}G\,\log s\int d^2\tr q {\delta\alpha
(\Q)\over\Q} s^{1-\alpha'\Q}e^{i\tr b\cdot\tr q}\eqno (4.1)$$
where, by eq. (3.7\a ) and (3.8), the singular $\Q$-dependence of
$\delta\alpha$ is of the form
$${\delta\alpha\over\Q}\simeq
-{G\over\pi}\log{\Q\over\lambda^2}-2{G\over\pi}\log{\Lambda\over\Q},
\eqno (4.2)$$
$\lambda $ ($\Lambda $) being IR (UV) cut-off. By replacing (4.2) into (4.1),
we obtain in the $\alpha'\to 0$ limit,
$$\delta_1^s(b,s)=-{2\over\pi}{G^2s\over b^2}\log s.\eqno (4.3)$$

This expression differs (by a factor of (${}-3$)) from the pure gravity
result (1.5).
It is natural to ascribe the above discrepancy to gravitino contributions, that
were computed by Lipatov [7] to be
$${\delta\alpha_N\over\Q}=NG\int{d^2k\over 4\pi^2}
{\Q-t_1-t_2\over t_1t_2}\eqno (4.4)$$
where N is the number of gravitinos. We see that the integrand in (4.4) is IR
finite, but yields an UV contribution
$$ {\delta\alpha_N\over\Q}=-{N\over 2}{G\over\pi}\log{\Lambda^2\over\Q}.
\eqno (4.5)$$
By summing graviton and gravitino contributions we obtain in general
$${\delta\alpha\over\Q}=-{G\over\pi}\log{\Q\over\lambda^2}+2{G\over\pi}\log
{\Lambda^2\over\Q}-{N\over 2}{G\over\pi}\log{\Lambda^2\over\Q},\eqno (4.6)$$
which for $N=0$ agrees with eq. (1.5) and for $N=8$ yields the SST
result (4.2).

A more direct way of arriving at the result (4.5) is to compute the
gravitino contribution to the one-loop absorptive part (fig. 2).
By using the graviton-gravitino amplitudes of ref. [12], we find (Appendix B)
$$ \Im {A_{g}(s,\Q)\over s}=N\pi G^2 s\log {\Lambda^2\over\Q}
P_{\uparrow\downarrow}=4N\int
d^2\tr b\,\Im\delta_{1g} (s,b)e^{i\tr b\cdot\tr q}\eqno (4.7)$$
where $P_{\uparrow\downarrow}$ denotes the projection operator on
s-channel states of opposite helicities.
It follows, by using the analyticity arguments of ACV II, that
the gravitino contributions to the one-loop eikonal are given by
$$\delta_1^{(N)}\equiv N\delta_{1g}={NG^2 s\over {\bf b}^2}
\biggl(iP_{\uparrow\downarrow}
-{1\over\pi}\log s\biggr)\,,\quad\quad\quad
{\delta\alpha_{N}\over\Q}=-{N\over 2}{G\over\pi}\log{\Lambda^2\over\Q},
\eqno (4.8)$$
in agreement with eq. (4.5).

We can now discuss the gravitino
contribution to the full scattering matrix at Planckian energies. We
look for a unitary S-matrix of eikonal type, with the factorized form
$$ S=S^{(G)}S^{(N)}\eqno (4.9)$$
where the graviton contribution
$$ S^{(G)}=\exp 2i(\delta_0+\delta_1+\delta_2)\times
({\rm inelastic\ terms})\eqno (4.10)$$
is the one found in ACV II and reported in eq. (1.4), while the
gravitino contribution is set to be the matrix
$$ S^{(N)}=e^{2i\delta_1^{(N)}}\left(1+2i\sqrt{\Im\delta_1^{(N)}}\,
\sigma_1 P_{\uparrow\downarrow}\right).\eqno (4.11)$$
Here $\delta_1^{(N)}$ is taken from eq. (4.8) and $\sigma_1$ is the
Pauli matrix whose entries represent the graviton and gravitino
channels.

Eq. (4.11) has been verified so far at one-loop level, but
can be argued to be valid at two loop level, by the following
analyticity and unitarity arguments.

First by expanding (4.9) at two loops, we find the elastic amplitude
$$ {A^{(2)}\over s}=F.T.\biggl[-{2\over
3}\delta_0^3+2i\delta_0\delta_1+\delta_2+
2i\delta_0\delta_1^{(N)}\biggr],\eqno (4.12)$$
where $F.T.$ indicates the Fourier transform.
The latter satisfies analyticity requirements for the following
reasons:
\cv
(a) The three-body imaginary part related to $\Im \delta_2$
is the same as for pure gravity.
In fact the gravitino diagrams of fig. 3 lack one factor of $\log s$,
with respect to the H-diagram contribution.
\cv
(b) The real part is consistent with analyticity. We need to
show this fact only for the last term in eq. (4.12), the pure gravity
part being the same as in ACV II. The phase of the last term is
$$\delta A^{(2)}=-2s\delta_0\,\Im\delta_1^{(N)}\biggl[
P_{\uparrow\downarrow}+{i\over\pi}\log s\biggr]\eqno (4.13)$$
and thus agrees with a combination of the $s$-$u$ symmetric analytic
functions
$$ s^3\bigl[\log (-s)\bigr]^2+u^3\bigl[\log (-u)\bigr]^2\sim
s^3\biggl({1\over 2}+{i\over\pi}\log s\biggr),\eqno (4.14\a )$$
$$s^3P_{\uparrow\downarrow}+u^3P_{\uparrow\uparrow}\sim
s^3\Bigl(P_{\uparrow\downarrow}-{1\over 2}\Bigr).\eqno (4.14\b )$$

Finally, the inelastic term in (4.11) agrees with the explicit
evaluation of Appendix B, and makes (4.11) unitary at two loop level.

It is amusing to note that according to Eqs. (4.9) and (4.12) the spin
dependent part of the gravitino contributions to the superstring amplitude
up to two loops takes the form (N=8)
$$P_{\uparrow\downarrow} 4{\Delta\over s}
\left(i\delta_0^2\left( b\right)-{2\over 3}\delta_0^3\left( b\right)\right),
\quad\quad\quad \Delta\equiv \nabla_{b}^2,\eqno (4.15)$$
while the SST ( and N=8 Supergravity) kinematical factor has the asymptotic
form
$$K_{cl}\sim\Bigl(1+4{t\over s}P_{\uparrow\downarrow}\Bigr)
\Bigl({s\over2}\Bigr)^4.\eqno (4.16)$$
By replacing $t=-\Q$ by $\Delta $ in impact parameter space we see
the result (4.15) is consistent with factorization of the kinematical
factor $K_{cl}$ even at two loop level.

The above arguments suggest that the elastic two loop contribution
to the eikonal $\delta_2$ computed by ACV II is not modified
by gravitino contributions. This fact is perhaps not surprising, in
view of the semiclassical interpretation of $\Re\,\delta_2$, given
in eq. (1.5). The fermionic gravitino field appears not to contribute to
the long distance eikonal in the leading $(1/\hbar)$ limit $R\sim GE
\gg\lambda_P\sim \sqrt{G\hbar}$.

In conclusion, our SST calculation confirms that massless gravitino terms
do contribute to the one-loop quantum corrections to the eikonal, similarly to
what happens in the massive scattering case [13,14]. On the other hand, we have
also argued that no such contribution is present in the ``classical'' ACV
correction $\delta_2$, which becomes the most important one when the
gravitational radius exceeds the Planck length.
\bigskip
\noindent{{\bf Acknowledgments}}
One of us (A.B.) is grateful to Nordita and Niels Bohr
Institute for the warm hospitality.
\vfill\eject
\sez{Appendix A}
We describe in the following how to compute the trajectory
renormalization, in the limit of small transverse momenta, starting from
the integral expression given in eq. (3.4). First of all we observe that
$\epsilon=0$ and $A\to 0$ are the regions of interest to obtain $\Q=0$
singularities in four dimension.
Therefore we can use the following approximation
$$\eqalignno {
I_{{\rm pole}}=&{1\over \pi^2}
\left(\sigma\over 2\right)^{-1-\Q}\log \sigma
\int_{\gamma}{dn\over 2\pi i}{dm\over 2\pi i}
{\Gamma(-n)\Gamma(-m)\over \Gamma(n+1)\Gamma(m+1)}\cr
&\times\int_0^1 dx\,B^{-2}
\left(-{A^2\over 8}\right)&(\A .1)\cr}$$
where $A$ and $B$ are given in eqs. (3.3\b ) and (3.2\b ). By defining
$t_1,\ t_2$ in terms of the loop momenta as in eq. (3.7\b ),
the integration over $x$ is transformed by the Feynman integral representation
$$\int_0^1 dx {1\over \Q x(1-x)+2nx+2m(1-x)}=\int {d^2{\bf k}\over\pi}
{1\over (2n+t_1)(2m+t_2)}\eqno (\A .2)$$
and its derivative
$$\eqalignno{&\int_0^1 dx {-1\over \left[ \Q x(1-x)+2nx+2m(1-x)\right]^2}\cr
&\quad =\int {d^2{\bf k}\over\pi}
\left({\partial\over\partial {t_1}}+{\partial\over\partial {t_2}}\right)
{1\over (2n+t_1)(2m+t_2)}. &(\A .3)\cr}$$
Therefore we can reformulate the integral (A.1) in the following way
$$\eqalignno {
I_{{\rm pole}}&=\left(\sigma\over 2\right)^{-1-\Q}\log \sigma
\int {d^2{\bf k}\over(2\pi)^3}
\left({\partial\over\partial {t_1}}+{\partial\over\partial {t_2}}\right)
\int_{\gamma}{dn\over 2\pi i}{dm\over 2\pi i}
{\Gamma(-n)\Gamma(-m)\over \Gamma(n+1)\Gamma(m+1)}\cr
&\qquad\times{\left(2n+2m+\Q\right)^2\over (2n+t_1)(2m+t_2)}.&(\A .4)\cr}$$
We proceed now to integrate along the contour $\gamma$ in the $n$ and $m$
complex plane shown in fig. 4, by deforming it to the left as usual. We pick up
the poles in $2n=-t_1,\ 2m=-t_2$ because the singular $\Q\to 0$ behavior arises
from the pinching of the contour between them and the poles in $n=0,\ m=0$.
It is straightforward to obtain
$$I_{{\rm pole}}=\left(\sigma\over 2\right)^{-1-\Q}\log \sigma
\int {d^2{\bf k}\over(2\pi)^3}
\left({\partial\over\partial {t_1}}+{\partial\over\partial {t_2}}\right)
{(\Q-t_1-t_2)^2\over t_1 t_2}\eqno (\A .5)$$
and using the factors in eq. (2.7a) the expression  (3.7) follows.
\vfill\eject
\sez{Appendix B}
In the following we review an S-matrix computation of the leading term of
the gravitino box, in the regime of large center of mass energy
$s=4E_{\rm cm}^2$ and fixed momentum transfer $t=-\Q$. For semplicity we
consider $N=1$ Supergravity in $D=4$ and use amplitudes in the helicity basis,
explicitly given at tree level for gravitons and gravitinos in ref. [12]
$$ S=\delta_{fi}-i\left( 2\pi\right)^4\delta^4
\left( p_1 +p_2 -p_3 -p_4\right)\left( p_3,p_4;p_1,p_2\right)
\eqno (\C .1)$$
$$\eqalignno {
\A_{GG} \left( 2,\ \ 2;2,\ \ 2\right)&=8\pi G {s^4\over stu},\quad\quad\quad\,
\A_{Gg} \left( 2,\ \ 2;{3\over 2},\ \ {3\over 2}\right)=0\cr
\A_{GG} \left( 2,-2;2,-2\right)&=8\pi G {u^4\over stu},\quad\quad\quad\,
\A_{Gg} \left( 2,-2;{3\over 2},-{3\over 2}\right)=
- 8\pi G {u^2{\sqrt {tu}}\over ts}\cr
\A_{gg} \left( {3\over 2},{3\over 2};{3\over 2},{3\over 2}\right)&=8\pi G
{s^4\over stu},\quad\quad\quad
\A_{gg} \left( {3\over 2},-{3\over 2};{3\over 2},-{3\over 2}\right)
=8\pi G {u^4\over stu}&(\C .2)\cr}$$
In particular we notice that crossing simmetry requires for external
gravitons the following relation
$$ \A \left( 2,2;2,2; s,t,u\right)=\A \left( 2,-2;2,-2;u,t,s\right)$$
and that the supergravity kinematical factor, which coincide with the
superstring result at tree level, is given for external gravitons by
the following expression in the helicity base
$$ K\propto s^4\left(P_{\uparrow\uparrow\uparrow\uparrow}+
P_{\downarrow\downarrow\downarrow\downarrow}\right)+
u^4\left(P_{\uparrow\downarrow\uparrow\downarrow}+
P_{\downarrow\uparrow\downarrow\uparrow}\right)+
t^4\left(P_{\uparrow\downarrow\downarrow\uparrow}+
P_{\downarrow\uparrow\uparrow\downarrow}\right)\eqno (\C .3)$$
where the $P$'s are projection operators on external helicity states.
In the high energy limit the leading and next to leading terms conserve
the helicity of the fast legs and admit the simplified notation
$$K\simeq s^4+4s^3tP_{\uparrow\downarrow}\eqno (\C .4)$$
where $P_{\uparrow\downarrow}=
P_{\uparrow\downarrow\uparrow\downarrow}+
P_{\downarrow\uparrow\downarrow\uparrow}$ .

We compute now the s-channel discontinuity due to gravitinos according
to the asymptotic formula
$$ \Im \A^{+}_{GG}={1\over 16\pi^2 s}
\int d^2\k \A^{+}_{Gg}(\k )\A^{-}_{Gg}(\q -\k ).$$
It is straightforward to prove, using eq. (\C .2), that this is proportional
to $P_{\uparrow\downarrow}$, precisely
$$ \Im \A_{GG}={1\over 16\pi^2 s}(8\pi G)^2 s^3\int {d^2\k \over\sqrt {\k^2
\left( \q -\k\right)^2}}\cong 4\pi G^2 s^2 \log {\Lambda^2\over \Q}
P_{\uparrow\downarrow}. \eqno (\C .5)$$
The leading real part of this amplitude can be obtained observing that the
following $s$-$u$ symmetric analytic function has the same discontinuity
$$ s^2\log (-s)P_{\uparrow\downarrow} +
u^2\log (-u)P_{\uparrow\uparrow}\simeq s^2\left(-i\pi
P_{\uparrow\downarrow}+ \log s\right)\eqno (\C .6)$$
and the impact parameter transform (4.9) follows for $N$ supergravity.

Finally we remind that the projector in eq. (4.11) operates over gravitons and
gravitinos helicity states according to the chosen channel, and gives
the  following amplitude $\A_{Gg}$ for the production of any pair of gravitinos
$$\A_{Gg}=4s\int d^2\b e^{i\b \cdot \q}{\sqrt {\Im \delta_{1g}}}=
{8\pi Gs^{3\over 2} \over \vert\q\vert}
\eqno (\C .7)$$
which is easily verified to give the same result as eq. (B.2) in the high
energy limit.
\vfill\eject
\sez {References}
\vskip0.5cm
\refno {1} G. 't Hooft, Phys. Lett. B198 (1987) 61;
Nucl. Phys. B304 (1988) 867; Nucl. Phys. B335 (1990) 138.
\refno {2} I. Muzinich and M. Soldate, Phys. Rev. D 37 (1988) 353.
\refno {3} D. Amati, M. Ciafaloni and G. Veneziano, Phys. Lett. B197
(1987) 81; Int. J. Mod. Phys. 3A (1988) 1615, hereafter referred to as
ACV I.
\refno {4} D. Amati, M. Ciafaloni and G. Veneziano, Nucl. Phys. B347
(1990) 550, hereafter referred to as ACV II.
\refno {5} E. Verlinde and H. Verlinde, {\it Scattering at planckian energies},
Princeton preprint PUTP-1279 (1991); see also R.Kallosh, {\it Geometry of
scattering at Planckian energies}, Stanford preprint SU-ITP 903 (1991).
\refno {6} D. Amati, M. Ciafaloni and G. Veneziano, {\it Planckian scattering
beyond the semiclassical approximation}, CERN preprint TH.6395/92.
\refno {7} L. N. Lipatov, Phys. Lett B116 (1982) 411;
Nucl. Phys. B365 (1991) 614.
\refno {8} B. Sundborg, Nucl. Phys. B306 (1988) 545.
\refno {9} M. Ademollo, A. Bellini and M. Ciafaloni, Phys. Lett.
B223 (1989) 318; Nucl. Phys. B338 (1990) 114.
\refno {10} M. B. Green, J. H. Schwarz and E. Witten,
{\it Superstring theory} (Cambridge University Press, Cambridge, 1987).
\refno {11} A. Bellini, G. Cristofano, M. Fabbrichesi and K. Roland,
Nucl. Phys. B356 (1991) 69.
\refno {12} P. Van Nieuwenhuizen, Phys. Rep. 68 (1981) 169; M.T. Grisaru,
P. Van Nieuwenhuizen and H.N. Pendleton, Phys. Rev. 15D (1977) 996.
\refno {13} E. Gava, R. Iengo and C.-J. Zhu, Nucl. Phys. B323 (1989) 585.
\refno {14} G. Cristofano, M. Fabbrichesi and K. Roland, Phys. Lett. B246
(1990) 45.

\vfill\eject
\bye